\documentclass{scrartcl}

\usepackage{array,xspace,multirow,hhline,tikz,tabularx,booktabs,fixltx2e}

\usepackage[numbers]{natbib}
\usepackage{amsmath,amssymb,amsfonts,amsthm}

\newtheorem{definition}{Definition}%
\newtheorem{theorem}{Theorem}%
\newtheorem{lemma}{Lemma}%
\newtheorem{corollary}{Corollary}%
\newtheorem{example}{Example}%

\newcommand{\secref}[1]{Section~\ref{#1}}

\newcommand{\lemref}[1]{Lemma~\ref{#1}}
\newcommand{\thmref}[1]{Theorem~\ref{#1}}

\newcommand{\exref}[1]{Example~\ref{#1}}

\newcommand{\remref}[1]{Remark~\ref{#1}}

\newcommand{\ie}{i.e.,\xspace}
\newcommand{\eg}{e.g.,\xspace}

\usepackage[mathcal]{euscript}

\usepackage{verbatim,ifthen}
\usetikzlibrary {shapes}
\usepackage{accents}  %
\usepackage{enumitem}

\newcommand{\fone}{\mathcal{F}}

\newcommand{\teq}[1][]{\ifthenelse{\equal{#1}{}}{\mathit{TEQ}}{\mathit{TEQ}(#1)}}
\newcommand{\me}[1][]{\ifthenelse{\equal{#1}{}}{\mathit{ME}}{\mathit{ME}(#1)}}
\newcommand{\mc}[1][]{\ifthenelse{\equal{#1}{}}{\mathit{MC}}{\mathit{MC}(#1)}}
\newcommand{\uc}[1][]{\ifthenelse{\equal{#1}{}}{\mathit{UC}}{\mathit{UC}(#1)}}
\newcommand{\ba}[1][]{\ifthenelse{\equal{#1}{}}{\mathit{BA}}{\mathit{BA}(#1)}}
\newcommand{\cnl}[1][]{\ifthenelse{\equal{#1}{}}{\mathit{CNL}}{\mathit{CNL}(#1)}}
\newcommand{\tc}[1][]{\ifthenelse{\equal{#1}{}}{\mathit{TC}}{\mathit{TC}(#1)}}
\newcommand{\co}[1][]{\ifthenelse{\equal{#1}{}}{\mathit{CO}}{\mathit{CO}(#1)}}
\newcommand{\bp}[1][]{\ifthenelse{\equal{#1}{}}{\mathit{BP}}{\mathit{BP}(#1)}}

\newcommand{\teqrel}[1][]{\ifthenelse{\equal{#1}{}}{\boldsymbol\rightarrow}{\teqrel_{#1}}}

\renewcommand{\mathrel}[1]{\,#1\,}

\newcommand{\midd}{\mathrel{:}}

\newcounter{remark}
\newenvironment{remark}[1][]{\refstepcounter{remark} \ifthenelse{\equal{#1}{}}{\noindent\emph{Remark~\theremark. }}{\noindent \emph{Remark~\theremark~(#1).}}}{\medskip}

\title{Set-Rationalizable Choice and Self-Stability}

\author{Felix Brandt \quad and \quad Paul Harrenstein\\
Technische Universit\"at M\"unchen\\85748 Garching bei M\"unchen, Germany\\
\texttt{\small \{brandtf,harrenst\}@in.tum.de}}

\newcommand{\set}[1]{\{#1\}}

\newlength{\WORDWIDTH}

\date{}

\begin{document}

\maketitle

\begin{abstract}
Rationalizability and similar notions of consistency have proved to be highly problematic in the context of social choice, as witnessed by a range of impossibility results, among which Arrow's is the most prominent. We propose to rationalize choice functions by preference relations over sets of alternatives (\emph{set-rationalizability}) and introduce two consistency conditions,~$\widehat\alpha$ and~$\widehat\gamma$, which are defined in analogy to Sen's~$\alpha$ and~$\gamma$. We find that a choice function satisfies~$\widehat\alpha$ if and only if it is set-rationalizable and that it satisfies~$\widehat\alpha$ and~$\widehat\gamma$ if and only if it is \emph{self-stable}, a new concept based on earlier work by Dutta. The class of self-stable social choice functions contains a number of appealing Condorcet extensions.
\end{abstract}

\noindent\textbf{Keywords: }Choice Theory, Rationalizability, Consistency, Stable Sets, Social Choice Theory\\
\noindent\textbf{JEL Classifications Codes: }D01, D71

\section{Introduction}
\label{sec:intro}

Arguably the most basic model of individual and collective choice is a \emph{choice function}, which associates with each set~$A$ of feasible alternatives a non-empty subset~$S(A)\subseteq A$. 
Apparently, not every choice function complies with our intuitive understanding of rationality. Consider, for example, the choice function~$S$ with $S(\set{a,b})=\set a$ and $S(\set{a,b,c})=\set{b}$. Doubts as to an agent's rationality could be raised, if, when offered the choice between apple pie and brownies, he were to choose the former, but the latter, when told that cr{\`e}me caramel is also an option.%
\footnote{\citet{Sen93a,Sen97a} has argued against imposing internal consistency conditions on rational choice. His examples usually involve a kind of context-dependence, like a modest person choosing a particular piece of cake only if it is not the largest available. Usually this context-dependence can be avoided by redefining the alternatives. Thus, for the purposes of this paper the context-independence of choice is merely a modeling assumption.}
In microeconomic theory, the existence of a binary relation~$R$ on all alternatives such that~$S$ returns precisely the maximal elements according to~$R$ from any feasible set is a common rationality condition on choice functions. Choice functions for which this is the case are called \emph{rationalizable} \citep[see, \eg][]{Rich66a,Herz73a,BBKS76a,Moul85a}.\footnote{Rationalizable choice functions have also been referred to as \emph{binary} \citep{Schw76a}, \emph{normal} \citep{Sen77a}, and \emph{reasonable} \citep{Alli99a}.} Rationalizable choice functions have been characterized using two \emph{consistency} conditions that relate choices within feasible sets of variable size, namely conditions $\alpha$ and $\gamma$ \citep{Sen71a}.  Clearly, acyclicity of the strict part~$P$ of~$R$ is necessary and sufficient for~$S$ to be rationalizable if every finite set of alternatives is feasible.
Stronger rationality conditions can be obtained by requiring the rationalizing relation~$R$ to satisfy certain structural restrictions, such as completeness, transitivity, or quasi-transitivity (\ie transitivity of~$P$). 

The above considerations have had a profound impact on the theory of social choice, in particular on the interpretation of Arrow's general impossibility theorem \citep{Arro51a}, which states the impossibility of social choice functions (SCFs) that satisfy four intuitive criteria, including rationalizability via a transitive preference relation.
An obvious way around Arrow's disturbing result is to relax this condition, \eg by requiring SCFs to be merely rationalizable.
Although this approach does allow for some SCFs that also meet the remaining three criteria, these functions turned out to be highly objectionable, usually on grounds of involving a weak kind of dictatorship or violating other conditions deemed indispensable for rational social choice \citep[for an overview of the extensive literature, see ][]{BBKS76a,Kell78a,Schw86a,Sen77a,Sen86a,CaKe02a}.  \citet[][page 5]{Sen95a} concludes that
\begin{quotation}
	[\dots] the arbitrariness of power of which Arrow's case of dictatorship is an extreme example, lingers in one form or another even when transitivity is dropped, so long as \emph{some} regularity is demanded (such as the absence of cycles).\hfill 
\end{quotation}

One possibility to escape the haunting impossibility of rationalizable social choice is to require only~$\alpha$ or~$\gamma$ but not both at the same time. It turns out that $\alpha$ (and even substantially weakened versions of $\alpha$) give rise to impossibility results that retain Arrow's spirit \citep{Sen77a}. By contrast, there are a number of SCFs that satisfy~$\gamma$. The smallest one among these based on majority rule is the \emph{uncovered set} \citep{Fish77a,Mill80a,Moul86a}.

In this paper, we approach the matter from a slightly different angle. Choice functions are defined so as to select \emph{subsets} of alternatives from each feasible set, rather than a single alternative. Still, the consistency and rationality conditions on choice functions have been defined in terms of alternatives. Taking cue from this observation, we propose an alternative notion of rationality called \emph{set-rationalizability}. A choice function~$S$ is \emph{set-rationalizable} if a binary relation~$R$ on all non-empty subsets of alternatives can be found such that for each feasible subset~$A$, $S(A)$ is the unique maximal set with respect to~$R$ among all non-empty subsets of~$A$.

We find that set-rationalizable choice functions can be characterized by $\widehat\alpha$, a natural variant of~$\alpha$ defined in terms of sets rather than alternatives. Despite its intuitive appeal, $\widehat\alpha$ has played a remarkably small role in (social) choice theory \citep[][]{Cher54a,AiAl95a}. Yet, it differentiates quite a number of well-known choice functions. In particular, we show that various prominent SCFs---such as all scoring rules, all scoring runoff rules, and all weak Condorcet extensions---do {not} satisfy $\widehat\alpha$, whereas a handful of Condorcet extensions---such as weak closure maximality, the minimal covering set, and the bipartisan set---do.

For our second result, we introduce a new property~$\widehat\gamma$, which relates to~$\gamma$  as~$\widehat\alpha$ does to~$\alpha$. It turns out that~$\widehat\alpha$ and~$\widehat\gamma$ characterize the class of \emph{self-stable} choice functions, whose definition is inspired by earlier work of~\citet{Dutt88a} and~\citet{Bran08a}. Despite the logical independence of $\widehat\alpha$ and $\widehat\gamma$, the class of self-stable SCFs also contains the Condorcet extensions mentioned above. These SCFs furthermore satisfy all conditions typically appearing in Arrovian impossibility results except rationalizability, \ie $\alpha$ and $\gamma$. Accordingly, by replacing~$\alpha$ and~$\gamma$ with~$\widehat\alpha$ and~$\widehat\gamma$, the impossibility of rationalizable social choice can be avoided and turned into a possibility result.

\section{Preliminaries}\label{sec:prelim}

Let~$U$ be a universe of %
\emph{alternatives}. 
Throughout this paper, we assume the set of \emph{feasible subsets} of~$U$ to be given by~$\fone(U)$, the set of finite and non-empty subsets of~$U$.  Our central object of study are \emph{choice functions}, \ie functions $S:\fone(U)\rightarrow \fone(U)$ such that $S(A)\subseteq A$ for all feasible sets~$A$.

A choice function~$S$ is called \emph{rationalizable} if there exists a binary relation~$R$ on~$U$ such that for each feasible set~$A$ and each alternative~$x\in A$,
\[
	\text{$x\in S(A)$ if and only if $y\mathrel Px$ for no~$y\in A$,}
\]
where~$P$ is the strict part of~$R$.
Observe that acyclicity of~$P$ is required to guarantee that~$S$ invariably returns a non-empty set.

Two typical candidates for the rationalizing relation are the \emph{base relation} $\overline R_S$ %
and the \emph{revealed preference relation} $R_S$, %
which, for all alternatives~$x$ and~$y$, are given by
\begin{gather*}
	\text{$x\mathrel{\overline R_S}y$ if and only if $x\in S(\set{x,y})$, and}\\
	\text{
	$x\mathrel{R_S}y$
	if and only if $x\in S(X)$ for some~$X$ with~$y\in X$.
	}
\end{gather*}
Thus, the revealed preference relation relates~$x$ to~$y$ if $x$ is chosen in the presence of~$y$ and possibly other alternatives, whereas the base relation only relates~$x$ to~$y$ if~$x$ is chosen in the exclusive presence of~$y$.

Rationalizable choice functions are characterized by a consistency axiom, which \citet{Schw76a} defined such that for all feasible sets~$A$ and~$B$ and all alternatives~$x\in A\cap B$, %
\[
\text{
$x\in S(A\cup B)$
if and only if 
$x\in S(A)$ and $x\in S(B)$.
}
\]
The above equivalence can be factorized into two implications, viz. the conditions~$\alpha$ and~$\gamma$ \citep{Sen71a} for feasible sets~$A$ and~$B$ and alternatives~$x\in A\cap B$,\footnote{The definitions of $\alpha$ and $\gamma$ given here are equivalent, but not syntactically identical, to Sen's original ones. They are chosen so as they reveal their similarity to $\widehat\alpha$ and $\widehat\gamma$ below.}
\settowidth{\WORDWIDTH}{$\alpha$} 
\begin{gather}
	\tag{\text{$\alpha$}}
	\text{
		if 
		$x\in S(A\cup B)$
		then
		$x\in S(A)$ and $x\in S(B)$,
		}\\[1ex]
	\tag{\makebox[\WORDWIDTH][c]{$\gamma$}}
	\text{
		if
		$x\in S(A)$ and $x\in S(B)$
		then
		$x\in S(A\cup B)$.
		}
\end{gather}
Axiom~$\alpha$ is a \emph{contraction} consistency property, which states that alternatives that are chosen in a feasible set are still chosen in feasible subsets. By contrast,~$\gamma$ is an \emph{expansion} consistency property, which states that alternatives chosen in two feasible sets are also chosen in their union. \citet{Sen71a} proved that a choice function~$S$ is rationalizable if and only if it satisfies both~$\alpha$ and~$\gamma$, with the witnessing relations~$\overline R_S$ and $R_S$, which are identical in the presence of~$\alpha$.

\begin{theorem}[\citeauthor{Sen71a}, \citeyear{Sen71a}]\label{thm:Sen71}
	A choice function is rationalizable if and only if it satisfies both~$\alpha$ and~$\gamma$. 
\end{theorem}

Similar results can also be obtained if stronger requirements are imposed on the rationalizing relation \citep[see, \eg][]{Sen77a,Moul85a,Schw76a}. For instance, \citet{Arro59a} showed that a choice function can be rationalized by a complete and transitive relation if and only if it satisfies the \emph{weak axiom of revealed preference (WARP)}---a consistency condition, first proposed by \citet{Samu38a}, which is stronger than the conjunction of~$\alpha$ and~$\gamma$ and central to large parts of microeconomic theory. 
There is a range of results stating the impossibility of SCFs satisfying weaker versions of WARP in a satisfactory way \citep[see, \eg][]{Kell78a,Schw86a,CaKe02a,Bank95a}. Among these, the results by \citet{MCSo72a} and \citet{BlDe77a} deserve special mention as they concern rationalizability instead of WARP. 
For further characterizations of rationalizable social choice the reader is referred to \citet{Moul85b}, \citet{Bank95a}, and \citet{AuBa00a}.

\section{Set-Rationalizable Choice}

In analogy to the definitions of \secref{sec:prelim}, we now define the concept of set-rationalizability along with the base and revealed preference relations over sets of alternatives, and properties~$\widehat\alpha$ and $\widehat\gamma$. The main result of this section is that set-rationalizable choice is completely characterized by~$\widehat\alpha$.

We say a choice function is \emph{set-rationalizable} if it can be rationalized via a preference relation on sets of alternatives.

\begin{definition}
A choice function~$S$ is \emph{set-rationalizable} if there exists a binary relation~$R\subseteq\fone(U)\times\fone(U)$ such that for all feasible sets~$A,X\in\fone(U)$,
\[
	\text{$X=S(A)$ if and only if $Y\mathrel PX$ for no~$Y\in\fone(A)$,}
\] 
where~$P$ is the strict part of~$R$.
\end{definition}

Observe that~$S$ is set-rationalizable only if for each feasible set~$A$, $S(A)$ is the \emph{unique} maximal feasible set~$X\in\fone(A)$ in~$R$. Also observe that we do not require the rationalizing relation to be acyclic.

We define the \emph{base relation $\overline R_S$} and the \emph{revealed preference relation~$\widehat R_S$} of a choice function~$S$ on sets as follows:\footnote{Given a choice function~$S$, the base relation on sets is a natural extension of the base relation on alternatives and, hence, both are denoted by~$\overline R_S$.}
\begin{gather*}
\text{$A\mathrel{\overline R_S} B$ if and only if $A=S(A\cup B)$,}\\
\text{$A\mathrel{\widehat R_S} B$ if and only if $A=S(X)$ for some $X$ with $B\subseteq X$.}
\end{gather*}

Condition $\widehat\alpha$ is defined as a natural variant of~$\alpha$ that makes reference to the entire set of chosen alternatives rather than its individual elements.
\begin{definition}
A choice function~$S$ satisfies $\widehat\alpha$, if for all feasible sets~$A$, $B$, and~$X$ with $X\subseteq A\cap B$,
\[
\tag{$\widehat \alpha$}
	\text{
	    if
		$X=S(A\cup B)$ 
		then
		$X=S(A)$ and $X=S(B)$\text.
	}
\]
\end{definition}
The standard contraction consistency condition~$\alpha$ is logically independent from~$\widehat\alpha$: the former does not imply the latter, nor the latter the former (see \exref{ex:setrevrel2}).
Moreover, $\widehat\alpha$ is not a contraction consistency property according to Sen's original terminology \citep[see, \eg][]{Sen77a}. It does not only require that chosen alternatives remain in the choice set when the feasible set is reduced, but also that unchosen alternatives remain outside the choice set. Thus, it has the flavor of both contraction and expansion consistency (see Remarks~\ref{rem:sen} and~\ref{rem:pi}). 

In this paper, however, we are concerned with the choice set as a whole and
$\widehat\alpha$ merely says that the set~$S(A)$ chosen from a feasible set~$A$ is also chosen from any subset~$B$ of~$A$, provided the former contains~$S(A)$. This reading is reflected by the useful characterization of $\widehat\alpha$ given in the following lemma, which reveals that $\widehat\alpha$ is equivalent to an established condition known as \citeauthor{Cher54a}'s \emph{postulate $5^*$} \citep{Cher54a}, the \emph{strong superset property} \citep{Bord79a}, or \emph{outcast}~\citep{AiAl95a}.
\begin{lemma}\label{lemma:setSSP}
	A choice function~$S$ satisfies~$\widehat\alpha$ if and only if for all feasible sets $A$ and $B$,
	\[
		\text{if $S(A)\subseteq B\subseteq A$ then $S(A)=S(B)$.}
	\]
\end{lemma}
\begin{proof}
For the direction from left to right, let $S(A)\subseteq B\subseteq A$. Then, both $A\cup B=A$ and $B=A\cap B$. Hence, $S(A\cup B)=S(A)\subseteq B=A\cap B\text.$ Since $S$ satisfies~$\widehat\alpha$, $S(A)=S(B)$.

For the opposite direction, assume for an arbitrary non-empty set~$X$, both $X\subseteq A\cap B$ and $X=S(A\cup B)$. Then, obviously, both $S(A\cup B)\subseteq A\subseteq A\cup B$ and $S(A\cup B)\subseteq B\subseteq A\cup B$. It follows that $S(A\cup B)=S(A)$ and $S(A\cup B)=S(B)$.
\end{proof} 
As a corollary of \lemref{lemma:setSSP}, we have that choice functions~$S$ satisfying~$\widehat\alpha$, like those satisfying~$\alpha$, are \emph{idempotent}, \ie $S(S(A))=S(A)$ for all feasible sets~$A$.%

We define $\widehat\gamma$ in analogy to~$\gamma$ as follows.
\begin{definition}
A choice function~$S$ satisfies $\widehat\gamma$ if for all feasible sets~$A$, $B$, and~$X$,
\[
\tag{$\widehat\gamma$} 
	\text{ 
	    if
		$X=S(A)$ and $X=S(B)$
		then
		$X=S(A\cup B)$\text.
	}
\]
\end{definition} 
Thus, a choice function satisfies~$\widehat\gamma$, if, whenever it chooses $X$ from two different sets, it also chooses $X$ from their union.
As in the case of $\alpha$ and $\widehat{\alpha}$, $\gamma$ and $\widehat{\gamma}$ are logically independent.
However, $\widehat \gamma$ is implied by the conjunction of~$\alpha$ and~$\gamma$ (see \remref{rem:properties}).

Condition~$\widehat\gamma$ is reminiscent of the generalized Condorcet condition \citep[see, \eg][]{BBKS76a}, which requires that for all feasible sets~$A$ and all $a\in A$,
\[
	\text{if $S(\{a,b\})=\set a$ for all $b\in A$ then $S(A)=\set a$.}
\]
Choice functions that satisfy this condition we will refer to as \emph{generalized Condorcet extensions}. It is easily appreciated that $\widehat\gamma$ implies the generalized Condorcet condition. In the setting of social choice, \emph{Condorcet extensions} are commonly understood to be SCFs for which additionally choice over pairs is determined by majority rule.

As in the case of $\alpha$ and $\gamma$, a single intuitive consistency condition summarizes the conjunction of $\widehat\alpha$ and $\widehat\gamma$: for all feasible sets $A$, $B$, and~$X$ with $X\subseteq A\cap B$,
\[
	\text{ 
	    $X=S(A)$ and $X=S(B)$
		 if and only if
		$X=S(A\cup B)$\text.
	}
\]

For illustrative purposes, consider the following two examples.

\begin{example}\label{ex:setrevrel}
Let the choice function~$S$ over the universe~$\set{a,b,c}$ be given by the following table.

\begin{center}
\begin{minipage}{7em}
$\begin{array}{c}
	\begin{array}[b]{ll}
		X	&	S(X)\\\midrule
		\set{a,b}	& \set{a}\\
		\set{b,c}	&\set{b}\\
		\set{a,c}	&\set{c}\\
		\set{a,b,c}	&\set{a,b,c}\\
	\end{array}
\end{array}$
\end{minipage}
\hspace{2em}
\begin{minipage}{20em}\centering
\begin{tikzpicture}[xscale=.7,yscale=1.5]

	\tikzstyle{every node}=[inner sep=1pt]

	\draw (0,0) 	node(a){\set{a}};
	\draw (3,0) 	node(b){\set{b}};
	\draw (6,0) 	node(c){\set{c}};

	\draw (0,1) 	node(ab){\set{a,b}};
	\draw (6,1) 	node(ac){\set{a,c}};
	\draw (3,1) 	node(bc){\set{b,c}};

	\draw (3,2) 	node(abc){\set{a,b,c}};

	\draw[latex-] (a) ..controls ++(2,-.75) and ++(-2,-.75)..  (c);
	\draw[-latex] (a) -- (b);
	\draw[-latex] (a) -- (ab);

	\draw[-latex] (b) -- (c); 		
	\draw[-latex] (b) --  (bc);
		
	\draw[-latex] (c) -- (ac);

	\draw[-latex] (abc) ..controls ++(-1.7,-.5) and ++(.25,.25).. (a);			
	\draw[-latex] (abc) ..controls ++(-1,-.5) and ++(-1,.5).. (b);			
	\draw[-latex] (abc) ..controls ++(1.5,-.5) and ++(-.25,.25).. (c);			
	\draw[-latex] (abc) ..controls ++(-1.5,-.25) and ++(.5,.35).. (ab);			
	\draw[-latex] (abc) ..controls ++(1.5,-.25) and ++(-.5,.35).. (ac);			
	\draw[-latex] (abc) -- (bc);		
		
	\draw[-latex] (a) ..controls ++(-.6,-.6) and ++(.6,-.6).. (a);			
	\draw[-latex] (b) ..controls ++(-.6,-.6) and ++(.6,-.6).. (b);			
	\draw[-latex] (c) ..controls ++(-.6,-.6) and ++(.6,-.6).. (c);						
	\draw[-latex] (abc) ..controls ++(-.6,.6) and ++(.6,.6).. (abc);			
\end{tikzpicture}
\end{minipage}
\end{center}

The revealed preference relation on sets~$\widehat R_{S}$ and the base relation on sets~$\overline R_S$ coincide and are depicted in the graph on the right. A routine check reveals that~$S$ satisfies both~$\widehat\alpha$ and~$\widehat\gamma$ (while it fails to satisfy $\alpha$). Also observe that each feasible set~$X$ contains a subset that is maximal (with respect to~$\widehat R_S$) among the non-empty subsets of~$X$, \eg $\set{a,b,c}$ in~$\set{a,b,c}$ and $\set{a}$ in~$\set{a,b}$.
\end{example}

\begin{example}\label{ex:setrevrel2}
Let the choice function~$S$ over the universe~$\set{a,b,c}$ be given by the following table.

\begin{center}
	\begin{minipage}{7em}
	$\begin{array}{c}
		\begin{array}{ll}
			X	&	S(X)\\\midrule
			\set{a,b}	& \set{a,b}\\
			\set{b,c}	&\set{c}\\
			\set{a,c}	&\set{a}\\
			\set{a,b,c}	&\set{a}\\
		\end{array}
	\end{array}$
	\end{minipage}
	\hspace{2em}
	\begin{minipage}{20em}\centering
	\begin{tikzpicture}[xscale=.7,yscale=1.5]

		\tikzstyle{every node}=[inner sep=1pt]

		\draw (0,0) 	node(a){\set{a}};
		\draw (3,0) 	node(b){\set{b}};
		\draw (6,0) 	node(c){\set{c}};

		\draw (0,1) 	node(ab){\set{a,b}};
		\draw (6,1) 	node(ac){\set{a,c}};
		\draw (3,1) 	node(bc){\set{b,c}};

		\draw (3,2.1) 	node(abc){\set{a,b,c}};

		\draw[-latex] (a) -- (b);

		\draw[-latex] (a) ..controls ++(2,-.75) and ++(-2,-.75)..  (c);
		\draw[-latex] (a) .. controls ++(-.85,.5) and ++(-.35,-.25).. (ab);
		\draw[-latex] (a) ..controls ++(-4.4,.9) and ++(-7,2)..  (ac);
		\draw[-latex] (a) ..controls ++(-3.15,1.2) and ++(-3,1.2)..  (bc);
		\draw[-latex] (a) ..controls ++(-4,.3) and ++(-6.5,.4).. (abc);

		\draw[-latex] (c) -- (b); 		

		\draw[latex-] (bc) -- (c);

		\draw[-latex] (ab) ..controls ++(-.6,.6) and ++(.6,.6).. (ab);
		\draw[-latex] (ab) .. controls ++(.85,-.5) and ++( .35,.25).. (a);

		\draw[-latex] (ab) --  (b);

		\draw[-latex] (a) ..controls ++(-.6,-.6) and ++(.6,-.6).. (a);			
		\draw[-latex] (b) ..controls ++(-.6,-.6) and ++(.6,-.6).. (b);			
		\draw[-latex] (c) ..controls ++(-.6,-.6) and ++(.6,-.6).. (c);						
	\end{tikzpicture}
	\end{minipage}
\end{center}
	
$S$ is rationalizable via the relation given by $a\mathrel{P}c\mathrel{P}b$ and $a\mathrel{I}b$. Nevertheless, the revealed preference relation over sets, as depicted on the right, does not set-rationalize this choice function. Observe that both $\set a$ and $\set{a,b}$ are maximal in~$\set{a,b}$ with respect to the strict part of~$\widehat R_S$. As $S(\set{a,b,c})=\set{a}$ and $S(\set{a,b})=\set{a,b}$,~$S$ clearly does not satisfy~$\widehat\alpha$. 
Thus, the example proves that $\widehat \alpha$ is not a weakening of $\alpha$ (and not even of the conjunction of $\alpha$ and $\gamma$).
\end{example}

The first example shows that set-rationalizing relations need not be acyclic or complete. However, complete set-rationalizing relations can easily be obtained by adding indifferences between all pairs of incomparable alternatives.
By definition, $\overline R_S$ of any choice function $S$ is 
anti-symmetric, \ie $X\mathrel{\overline R_S}Y$ and $Y\mathrel{\overline R_S}X$ imply $X=Y$. In the presence of~$\widehat\alpha$,~$\widehat R_S$ and~$\overline R_S$ coincide and are thus both anti-symmetric.

Set-rationalizable choice functions are characterized by~$\widehat\alpha$.\footnote{\citeauthor{Moul85a} shows a similar statement for single-valued choice functions~\citep{Moul85a}.}
\begin{theorem}\label{thm:SetRatiffhatalphahatgamma}
A choice function is set-rationalizable if and only if it satisfies~$\widehat\alpha$.
\end{theorem}

\begin{proof}
For the direction from left to right, assume~$S$ is set-rationalizable and let $\mathrel R$ be the witnessing binary relation on sets. Now consider arbitrary feasible sets~$A$, $B$ and~$X$ with $X\subseteq A\cap B$ and assume that $X=S(A\cup B)$. As~$R$ set-rationalizes~$S$, we have $Y\mathrel P X$ for no $Y\subseteq A\cup B$. Accordingly, there is no $Y\subseteq A$ such that $Y\mathrel P S(A\cup B)$ either. By definition of set-rationalizability it follows that $S(A)=S(A\cup B)$. By an analogous argument, we also obtain $S(B)=S(A\cup B)$, as desired.

For the opposite direction, assume~$S$ to satisfy~$\widehat\alpha$ and consider an arbitrary feasible set~$A$ and an arbitrary $Y\in\fone(A)\setminus S(A)$. Then, $S(A)\subseteq S(A)\cup Y\subseteq A$. In virtue of \lemref{lemma:setSSP} it follows that $S(A)=S(S(A)\cup Y)$. 
By definition of~$\overline R_S$, then $S(A)\mathrel{\overline R_S}Y$. Moreover, due to the anti-symmetry of~$\overline R_S$, we have $S(A)\mathrel{\overline P_S}Y$. 
We may conclude that $\overline R_S$ set-rationalizes~$S$.
 A similar argument holds for~$\widehat R_S$, which coincides with $\overline R_S$ in the presence of $\widehat \alpha$.
\end{proof}

In the proof of \thmref{thm:SetRatiffhatalphahatgamma}, it is the base and revealed preference relations on sets that are witness to the fact that choice functions satisfying~$\widehat\alpha$ are set-rationalizable. In contrast to \citeauthor{Sen71a}'s \thmref{thm:Sen71}, however, the base and revealed preference relations on sets are not the unique relations that can achieve this. It is also worth observing that the proof shows that for each feasible set~$X$ and choice function~$S$ satisfying~$\widehat\alpha$, the selected set $S(X)$ is not merely a maximal set but also the unique \emph{maximum} set within~$X$ given~$\widehat R_S$, \ie $S(X)\mathrel{\widehat R_S} Y$ for all non-empty subsets~$Y$ of~$X$. 

\section{Self-Stability}

The importance of maximal---\ie undominated---alternatives stems from the fact that dominated alternatives can be upset by other alternatives; they are unstable. The rationale behind stable sets, as introduced by \citet{vNM44a}, is that this instability is only meaningful if an alternative is upset by something which itself is stable. Hence, a set of alternatives $X$ is said to be stable if it consists precisely of those alternatives not upset by $X$. In von Neumann and Morgenstern's original definition, $a$ is upset by $X$ if there exists some $b\in X$ such that $a\not\in S(\{a,b\})$ for some choice function $S$.
It turns out that the set consistency conditions introduced in the previous section bear a strong relationship to a notion of stability, where $a$ is upset by $X$ if $a\not\in S(X\cup \{a\})$ \citep[see also][]{Bran08a}. The stability of choice sets can then be formally defined as follows.
\begin{definition}
Let $A,X$ be feasible sets and $S$ a choice function. $X$ is \emph{$S$-stable in~$A$} if 
	\[
		X=\set{a\in A\midd a\in S(X\cup\set a)}\text.
	\]
\end{definition}
Equivalently,~$X$ is $S$-stable in $A$ if it satisfies both \emph{internal} and \emph{external $S$-stability}:
\begin{gather}
\tag{internal $S$-stability}
S(X)=X \text,
\\[1ex]
\tag{external $S$-stability}
a\not\in S(X\cup\{a\}) \text{ for all } a\in A\setminus X\text.
\end{gather}
The intuition underlying this formulation is that there should be no reason to restrict the selection by excluding some alternative from it and, secondly, there should be an argument against each proposal to include an outside alternative into the selection.

For some choice functions~$S$, a unique inclusion-minimal $S$-stable set generally exists. If that is the case, we use~$\widehat S$ to denote the choice function that returns the unique minimal $S$-stable set in each feasible set and say that $\widehat S$ is \emph{well-defined}. Within the setting of social choice, a prominent example is \citeauthor{Dutt88a}'s \emph{minimal covering set $\mc$}~\citep{Dutt88a,DuLa99a}, which is defined as 
$\mc=\widehat\uc$, where $\uc$ is the uncovered set \citep{Fish77a,Mill80a}. Proving that a choice function~$\widehat S$ is well-defined frequently turns out to be highly non-trivial~\citep{Bran08a}.

We find that there is a close connection between~$\widehat\gamma$ and minimal $S$-stable sets.
\begin{lemma}\label{lem:hatplusuniqueimphatgamma}
	Let $S$ be a choice function such that $\widehat S$ is well-defined. Then $\widehat S$ satisfies~$\widehat\gamma$.
\end{lemma}
\begin{proof}
Consider arbitrary feasible sets~$A,B,X$ and assume that $\widehat S(A)=\widehat S(B)=X$. Trivially, as~$X$ is internally $S$-stable in~$A$, so is~$X$ in~$A\cup B$. To appreciate that~$X$ is also externally $S$-stable in~$A\cup B$, consider an arbitrary~$x\in(A\cup B)\setminus X$. Then, $x\in A\setminus X$ or $x\in B\setminus X$. In either case, $x\notin S(X\cup\set{x})$, by external $S$-stability of~$X$ in~$A$ if the former, and by external $S$-stability of~$X$ in~$B$ if the latter. Also observe that any subset of~$X$ that is $S$-stable in~$A\cup B$ would also have been $S$-stable in both~$A$ and~$B$. Hence,~$X$ is minimal $S$-stable in~$A\cup B$. Having assumed that~$\widehat S$ is well-defined, we may conclude that~$\widehat S(A\cup B)=X$.
\end{proof}

We now introduce the notion of self-stability. A choice function~$S$ is said to be {self-stable} if for each feasible set $A$, $S(A)$ is the unique (minimal) $S$-stable set in~$A$.
\begin{definition} 
	A choice function~$S$ is \emph{self-stable} if $\widehat S$ is well-defined and $S=\widehat S$.
\end{definition}

The class of self-stable choice functions is characterized by the conjunction of $\widehat\alpha$ and~$\widehat{\gamma}$.
\begin{theorem}\label{thm:selfstabilityiffhatalphahatgamma}
A choice function is self-stable if and only if it satisfies both~$\widehat\alpha$ and~$\widehat\gamma$.
\end{theorem}
\begin{proof}	
For the direction from left to right, assume~$S$ to be self-stable. Then,~$\widehat S$ is well-defined and $\widehat S=S$. \lemref{lem:hatplusuniqueimphatgamma} implies that~$S$ satisfies~$\widehat \gamma$. 
For~$\widehat\alpha$, consider arbitrary feasible sets $A,B$ such that $S(A)\subseteq B\subseteq A$. By virtue of \lemref{lemma:setSSP}, it suffices to show that $S(B)=S(A)$. First, observe that $S(A)$, which is $S$-stable in $A$, is also $S$-stable in $B$ since internal and external stability straightforwardly carry over from $A$ to its subset~$B$. Next, we show 
that $S(B)$ is not only the \emph{minimal} $S$-stable set in~$B$, but even the \emph{only} $S$-stable set in~$B$. To appreciate this, consider an arbitrary feasible set~$X\subseteq B$ with $X\neq S(B)$ and assume for contradiction that~$X$ is $S$-stable. By definition of $S(B)$ as the unique inclusion-minimal $S$-stable set in~$B$, it follows that $S(B)\subset X$. As $S(B)$ is $S$-stable in~$B$, $S(B)$ is obviously also $S$-stable in~$X$. Self-stability of~$S$ then requires that~$S(X)$, the \emph{minimal} $S$-stable set in~$X$, has to be contained in $S(B)$. Hence, $S(X)\subseteq S(B)\subset X$ and, in particular, $S(X)\neq X$, which is at variance with the internal stability of~$S$. As a consequence, $S(B)=S(A)$.

For the other direction, assume~$S$ satisfies both~$\widehat\alpha$ and~$\widehat\gamma$ and consider an arbitrary feasible set~$A$. For each $a\in A$, we have $S(A)\subseteq S(A)\cup\set{a}\subseteq A$. 
By $\widehat \alpha$ and \lemref{lemma:setSSP}, $S(S(A)\cup\set a)=S(A)$, which yields both internal and external stability of $S(A)$.
Finally, to see that $\widehat S$ is well-defined, consider an arbitrary $S$-stable set~$X$ in~$A$ and let $A\setminus X=\set{a_1,\ldots,a_k}$.
First, we show that $S(X\cup\{a_i\})=X$ for all $i\in\{1,\dots,k\}$. External stability implies that $a_i\not\in S(X\cup\{a_i\})$. Hence, by $\widehat \alpha$ and \lemref{lemma:setSSP}, $S(X\cup\{a_i\})=S(X)$, which by internal stability is identical to $X$. Repeated application of $\widehat \gamma$ then yields $S(X\cup\set{a_1,\dots,a_k})=S(X)$, which concludes the proof.
\end{proof}

As an immediate consequence of \thmref{thm:selfstabilityiffhatalphahatgamma} and the observation that~$\widehat\gamma$ implies the generalized Condorcet condition, we have the following corollary.
\begin{corollary}\label{cor:condorcet}
	Every self-stable choice function is a generalized Condorcet extension.
\end{corollary}
\noindent

Within the setting of social choice, only few SCFs turn out to be self-stable (or set-rationalizable). For instance, all scoring rules, all scoring runoff rules, and all weak Condorcet extensions fail to satisfy $\widehat \alpha$ (see \remref{rem:scoring}). Nevertheless, there is a small number of SCFs that are self-stable. Among them are \emph{Pareto's rule}, 
the \emph{omninomination rule},\footnote{This SCF chooses all alternatives that are ranked first by at least one voter \citep[see, \eg][]{Tayl05a}}  
\emph{weak closure maximality} (also known as the \emph{top cycle}, \emph{GETCHA}, or the \emph{Smith set}, see, \eg \citealp{Bord76a}), the \emph{minimal covering set} \citep{Dutt88a,DuLa99a}, and the \emph{bipartisan set} \citep{LLL93b,Lasl00a}.\footnote{\citet{Bran08a} defines an infinite hierarchy of self-stable SCFs. If we assume an odd number of agents with linear preferences, the class of self-stable SCFs is also conjectured to contain the \emph{tournament equilibrium set} \citep{Schw90a} and the \emph{minimal extending set}. Whether this is indeed the case depends on a certain graph-theoretic conjecture \citep{LLL93a,Bran08a}.}
Well-known SCFs that satisfy only one of~$\widehat\alpha$ and $\widehat\gamma$ appear to be less common. Still, \emph{strong} closure maximality \citep{Schw72a} is an example of an SCF that satisfies~$\widehat\gamma$ but not~$\widehat\alpha$. By contrast, the iterated elimination of Condorcet losers satisfies~$\widehat\alpha$ but not~$\widehat\gamma$.

By weakening transitive rationalizability to set-rationalizability (see \remref{rem:qt}), we have thus shown that appealing SCFs that also satisfy the other Arrovian postulates do exist.

\section{Concluding Remarks}
\begin{remark}[Sen's expansion and contraction]\label{rem:sen}
Condition~$\widehat \alpha$ can be split into two conditions that fall into Sen's categories: an expansion condition known as $\epsilon^+$ \citep{Bord83a} or \emph{Aizerman} \citep{Moul86a}, which requires that $S(B)\subseteq S(A)$ for all $S(A)\subseteq B\subseteq A$, and a corresponding contraction condition. Similarly, $\widehat \gamma$ can be factorized into two conditions.
\end{remark}

\begin{remark}[Path independence]\label{rem:pi}
	An influential and natural consistency condition that also has the flavor of both contraction and expansion is \emph{path independence} \citep{Plot73a}, which is satisfied if $S(A\cup B)=S(S(A)\cup S(B))$ for all $A$ and $B$. \citet{AiMa81a} have shown that path independence is equivalent to the conjunction of $\alpha$ and $\epsilon^+$. Since $\alpha$ is the strongest contraction consistency property and implies the contraction part of $\widehat \alpha$, it turns out that an alternative characterization can be obtained: a choice function is path independent if and only if it satisfies $\alpha$ and $\widehat \alpha$. Furthermore, since path independence implies $\widehat \gamma$, a choice function is path independent if and only if it satisfies $\alpha$, $\widehat \alpha$, and $\widehat \gamma$.
\end{remark}

\begin{remark}[Quasi-transitive rationalizability]\label{rem:qt}
\citet{Schw76a} has shown that quasi-transitive rationalizability is equivalent to the conjunction of $\alpha$, $\gamma$, and $\epsilon^+$. It is thus stronger than path independence and also implies both~$\widehat \alpha$ and~$\widehat \gamma$. 
So does the even stronger WARP condition.
\end{remark}

	\begin{remark}[Rationalizability implies $\widehat\gamma$]\label{rem:properties}
Assume $S$ satisfies both~$\alpha$ and~$\gamma$ and consider feasible sets~$X$,~$A$, and~$B$ with $X=S(A)$ and $X=S(B)$. The inclusion of $X$ in $S(A\cup B)$ follows immediately from~$\gamma$. To appreciate that also $S(A\cup B)\subseteq X$, consider an arbitrary $x\notin X$ and assume for contradiction that $x\in S(A\cup B)$. Then, either $x\in A$ or $x\in B$. Without loss of generality, we may assume the former. Clearly, $x\in (A\cup B)\cap A$ and $\alpha$ now implies that $x\in S(A)$, a contradiction.	
\end{remark}

\begin{remark}[Closure of set relations]
	The revealed preference relation on sets~$\widehat R_S$ of any choice function~$S$ that satisfies~$\widehat\alpha$ is closed under intersection, \ie for all feasible sets~$X$, $Y$, and~$Z$ such that $Y\cap Z\neq\emptyset$,
$
			\text{$X\mathrel{\widehat R_S} Y$ and $X\mathrel{\widehat R_S} Z$ imply $X\mathrel{\widehat R_S} Y\cap Z$.}
$	
Similarly, $\widehat R_S$ of a choice function~$S$ that satisfies~$\widehat\gamma$ is closed under union,\footnote{This condition is also known as \emph{robustness} \citep{Arle03a,BBP04a}.} \ie for all feasible sets~$X$, $Y$, and~$Z$,
$
	\text{$X\mathrel{\widehat R_S} Y$ and $X\mathrel{\widehat R_S} Z$ imply $X\mathrel{\widehat R_S} Y\cup Z$.}
$
\end{remark}

\begin{remark}[Scoring rules and weak Condorcet extensions]\label{rem:scoring}
The following preference profile (figures indicate numbers of agents) shows that many common SCFs do not satisfy~$\widehat \alpha$.

\[
		\begin{array}{ccc}
			3	&	2	&	1\\\midrule
			a	&	b	&	c\\
			c	&	a	&	b\\
			b	&	c	&	a
		\end{array}
\]

For all scoring rules (\eg plurality rule or Borda's rule), all scoring runoff rules (\eg Hare's rule or Coombs' rule), all weak Condorcet extensions---\ie SCFs that exclusively return the set of weak Condorcet winners whenever this set is non-empty---as well as a number of other common SCFs (\eg Kemeny's rule, Dodgson's rule, and Nanson's rule), the choice function for this profile is as in \exref{ex:setrevrel2} and therefore does not satisfy~$\widehat \alpha$.
\end{remark}

\section*{Acknowledgements}
We thank Nicholas Baigent, Cristopher Tyson, and an anonymous reviewer for helpful suggestions and Sean Horan for pointing out an error in a previous version of Remark~\ref{rem:qt}.
This material is based upon work supported by the Deutsche Forschungsgemeinschaft under grants BR~2312/3-3 and BR~2312/7-1.

\end{document}